\documentclass[lettersize,journal]{IEEEtran}
\usepackage{amsmath,amsfonts}
\usepackage{algorithmic}
\usepackage{algorithm}
\usepackage{array}
\usepackage[caption=false,font=footnotesize,labelfont=rm,textfont=rm]{subfig}
\usepackage{textcomp}
\usepackage{stfloats}
\usepackage{url}
\usepackage{verbatim}
\usepackage{graphicx}
\usepackage{cite}
\usepackage{multirow} 
\begin{document}

\title{Spatially Covariant Image Registration with Text Prompts}

\author{Xiang Chen, Min Liu, Rongguang Wang, Renjiu Hu, Dongdong Liu, Gaolei Li, and Hang Zhang
\thanks{Manuscript received April 19, 2021. This work was supported in part by the National Key Research and Development Program of China under Grant 2022YFE0134700, and in part by the National Natural Science Foundation of China under
Grant U22B2050, in part by the Science and Technology Program of
Changsha under Grant kq2102009. (Corresponding author: Hang Zhang.)}
\thanks{Xiang Chen, Min Liu are with the College of Electrical and Information Engineering, Hunan University, Changsha, China. (e-mail: xiangc@hnu.edu.cn, liu\_min@hnu.edu.cn).}
\thanks{Hang Zhang and Renjiu Hu are with Cornell University, New York, USA. (e-mail: hz459@cornell.edu, rh656@cornell.edu).}
\thanks{Rongguang Wang is with the University of Pennsylvania, Pennsylvania, USA. Dongdong Liu is with New York University, New York, USA. Gaolei Li is with Shanghai Jiao Tong University, Shanghai, China. (e-mail: rgw@seas.upenn.edu, ddliu@nyu.edu, gaolei\_li@sjtu.edu.cn).}}

\markboth{Journal of \LaTeX\ Class Files,~Vol.~14, No.~8, August~2021}%
{Shell \MakeLowercase{\textit{et al.}}: Spatially Covariant Medical Image Registration with CLIP Embedding}

\IEEEpubid{0000--0000/00\$00.00~\copyright~2021 IEEE}

\maketitle

\begin{abstract}
Medical images are often characterized by their structured anatomical representations and spatially inhomogeneous contrasts. Leveraging anatomical priors in neural networks can greatly enhance their utility in resource-constrained clinical settings. Prior research has harnessed such information for image segmentation, yet progress in deformable image registration has been modest. Our work introduces textSCF, a novel method that integrates spatially covariant filters and textual anatomical prompts encoded by visual-language models, to fill this gap. This approach optimizes an implicit function that correlates text embeddings of anatomical regions to filter weights. TextSCF not only boosts computational efficiency but can also retain or improve registration accuracy. By capturing the contextual interplay between anatomical regions, it offers impressive inter-regional transferability and the ability to preserve structural discontinuities during registration. TextSCF's performance has been rigorously tested on inter-subject brain MRI and abdominal CT registration tasks, outperforming existing state-of-the-art models in the MICCAI Learn2Reg 2021 challenge and leading the leaderboard. In abdominal registrations, textSCF's larger model variant improved the Dice score by 11.3\% over the second-best model, while its smaller variant maintained similar accuracy but with an 89.13\% reduction in network parameters and a 98.34\% decrease in computational operations.
\end{abstract}


\section{Introduction}
\label{sec:intro}
\IEEEPARstart{D}{eformable} image registration aims to find a dense, non-linear correspondence between a pair/group of images to estimate their alignment transformation.
This process is essential for numerous medical imaging applications, including tracking changes in longitudinal studies, measuring organ motion, and analyzing population-based studies \cite{viergever2016survey,haskins2020deep,chen2021deepSurvey}.
Traditional methods \cite{beg2005computing,ashburner2007fast,vercauteren2009diffeomorphic,avants2011reproducible,marstal2016simpleelastix} approach deformable image registration as a pairwise optimization problem.
This optimization typically requires numerous iterations to minimize the energy function, which is computationally expensive and consequently limits its application in real-time and large-scale volumetric image registration.
Additionally, these traditional methods depend on raw image contrasts, potentially leading to a loss of important contextual information.

\IEEEpubidadjcol 
Recent advances in convolutional neural networks (ConvNets) \cite{he2016deep,ronneberger2015u} and transformers \cite{liu2021swin} have changed the landscape of medical image registration research \cite{hering2022learn2reg}.
Voxelmorph \cite{balakrishnan2018unsupervised,balakrishnan2019voxelmorph} has enabled unsupervised learning and real-time registration.
Subsequent research has sought to enhance registration precision further by harnessing contextual cues from neural networks.
This can be seen in strategies such as leveraging auxiliary segmentation masks for anatomy awareness \cite{balakrishnan2019voxelmorph} or preserving discontinuities \cite{ng2020unsupervised,chen2021deepDiscontinuity}, employing cascaded \cite{zhao2019recursive,mok2020large}, parallel \cite{jia2022u}, or band-limited architectures \cite{wang2020deepflash,jia2023fourier}, as well as attention-based transformers \cite{chen2022transmorph,shi2022xmorpher,mok2022affine} to refine representation learning.
Despite their advances, these methods often underperform on datasets with large deformations or limited training instances.
While the integration of pre-trained anatomical embedding networks \cite{yan2022sam} with energy-based optimization presents a potential strategy to address these limitations \cite{liu2021same,li2023samconvex}, they generally do not attain the efficiency levels of learning-based approaches.

Typically, medical images obtained from modalities such as magnetic resonance imaging (MRI) and computerized tomography (CT) often exhibit structured anatomical patterns \cite{zhang2021all} and present spatially inhomogeneous contrasts \cite{zhang2023deda,zhang2022qsmrim}, lending themselves well for analysis using neural networks \cite{zhang2023deda}.
Acknowledging these characteristics leads us to two pivotal questions whose answers could mitigate the aforementioned challenges:
\textbf{Q1:} How can we leverage the innate prior information within each dataset, such as the consistent anatomical structures observed across different scans?
\textbf{Q2:} How might we incorporate external prior knowledge, like that from models pre-trained on disparate datasets?

To tackle the above challenges, we introduce a novel weakly-supervised learning framework that integrates the capabilities of large-scale visual-language model with spatially covariant filters (textSCF).
Our approach differs from previous methods by not only utilizing auxiliary segmentation masks in our loss function but also encoding these masks into an $N\times C$ embedding matrix, where $N$ is the number of labels and $C$ denotes the encoding length.
Building upon earlier work in spatially covariant lesion segmentation \cite{zhang2023spatially}, our framework optimizes an implicit function that associates this embedding matrix with corresponding filter weights, ensuring that the output is inherently spatially variant and aligned with anatomical regions.
Furthermore, we leverage the Contrastive Language-Image Pre-Training (CLIP) model \cite{radford2021learning} to generate the embedding matrix. 
By inputting text descriptions of anatomical regions, the model captures rich contextual information, such as the latent correlation of certain anatomical regions and discontinuities across these regions.

In this study, we put our method to the test on two different tasks: inter-subject brain registration via MRI and abdomen registration using CT scans. 
Our research has yielded several interesting findings:
\begin{itemize}
    \item Utilizing spatially covariant filters, textSCF demonstrated remarkable efficiency. 
    A scaled-down version of our model delivered comparable accuracy in abdomen registration while substantially cutting down on network parameters by 89.13\% and computational operations by 98.34\%.
    \item This research is the first to integrate text embeddings from visual-language model into volumetric image registration, improving the model's ability to contextually interpret anatomical structures.
    \item TextSCF can predict discontinuity-preserving deformation fields, capturing more realistic motion across multi-organs in real-world scenarios, which is generally ignored in previous registration methods.
    \item The textSCF approach consistently outperformed other leading methods in both brain and abdomen registration tasks. 
    A significant achievement was securing the first place on the MICCAI Learn2Reg 2021 challenge leaderboard at the time of our submission.
\end{itemize}

\section{Related Work}
\subsection{Learning-based Medical Image Registration}
Recent advancements in unsupervised ConvNets have significantly enhanced medical image registration, removing the dependence on ground-truth deformation fields, and achieving real-time performance.
VoxelMorph \cite{balakrishnan2018unsupervised, balakrishnan2019voxelmorph} pioneers the field by demonstrating that a generalized representation learned from a collection of image pairs could yield registration results on par with iterative approaches, and notably, in under a second for volumetric brain scans.
Further research has explored various network designs, including parallel \cite{kang2022dual, jia2022u} and cascaded architectures \cite{zhao2019recursive, mok2020large, jia2021learning}, as well as the incorporation of attention-based transformers \cite{chen2022transmorph} to refine the representation learning. 

In this paper, our focus is directed towards anatomically-aware methodologies \cite{balakrishnan2019voxelmorph,hering2021cnn,mok2021large,chen2021deepDiscontinuity} and the employment of pre-trained models from external datasets \cite{siebert2021fast,liu2021same,yan2022sam,li2023samconvex}, as they present promising solutions for enhancing registration accuracy, particularly for challenging datasets characterized by large deformations or limited training instances. 
Balakrishnan et al.~\cite{balakrishnan2019voxelmorph} and Mok et al.~\cite{mok2021large} apply auxiliary segmentation masks to regularize the training process using a Dice loss. 
Hering et al.~\cite{hering2021cnn} goes a step further by incorporating multiple anatomical constraints, such as a curvature regularizer and an anatomical keypoints loss. 
Chen et al.~\cite{chen2021deepDiscontinuity} utilizes segmentation masks to enforce deformation discontinuities between different organs, thereby preserving clinically important indices.
In terms of leveraging external data, Yan et al.~\cite{yan2022sam} introduces a self-supervised framework named SAM, which is capable of generating voxel-wise embeddings that provide semantically coherent features aiding in accurate registration. 
Building on SAM, SAME \cite{liu2021same} and SAMConvex \cite{li2023samconvex} exhibit improved registration performance on datasets with large deformations by incorporating more sophisticated network designs and combining energy-based optimization within ConvNets.
Different from conventional methods, our textSCF harnesses knowledge from both large-scale visual-language models \cite{radford2021learning} and specialized segmentation frameworks \cite{tang2022self,fischl2012freesurfer}.

\subsection{Spatipally Covariant Filters}
Translation invariance is a characteristic of ConvNets, yet it's been revealed that these networks can implicitly capture positional cues, presenting both opportunities and constraints.
Kayhan et al.~\cite{kayhan2020translation} reveals that zero padding implicitly leaks location information, where a stack of deeper convolutional layers can improve the position readout. 
On the other hand, Islam et al.~\cite{islam2019much} argues that removing such positional encodings could enforce stronger translation invariance, beneficial in certain classification tasks.
This suggests that ConvNets allocate a portion of their capacity to position encoding, implying that direct coordinate feeding as in CoordConv \cite{liu2018intriguing} can enhance network performance by utilizing its capacity.
Similarly, Elsayed et al.~\cite{elsayed2020revisiting} discovers that moderate relaxation of translation invariance could benefit classification tasks.
Zhang et al.~\cite{zhang2023spatially} introduces Spatially Covariant Pixel-aligned classifier (SCP), which relaxes translation invariance through an implicit function that maps image coordinates to classifier weights.
While SCP falls short for our registration task that needs precise per-pixel deformation vectors, we draw on its principles, using segmentation masks coupled with text prompts to direct the generation of spatially covariant filters.

\subsection{Text-driven Dense Predictions}
Large-scale visual-language pretraining models like CLIP \cite{radford2021learning} have shown impressive capabilities in complex visual tasks.
Despite its training on instance-level text-image pairs via contrastive learning, CLIP's efficacy extends to downstream tasks requiring per-pixel dense predictions. 
Its effectiveness is demonstrated in various applications such as semantic segmentation \cite{cheng2021per,zhou2022extract,luddecke2022image}, referent segmentation \cite{wang2022cris}, instance segmentation \cite{wang2022open,zhu2023segprompt}, and object detection \cite{gu2021open,rao2022denseclip,park2022per}.
Research leveraging these models for medical imaging, however, remains limited \cite{singhal2023large}. 
Most efforts concentrate on image instance-level analysis, like X-ray interpretation \cite{Wu_2023_ICCV,wang2022medclip}. 
Yet, Liu et al.~\cite{liu2023clip} has demonstrated that a tailored framework could harness anatomical relationships from text-image models for building universal segmentation models. 
Inspired by these advancements, our work pioneers the integration of text-image model CLIP \cite{radford2021learning} with spatially covariant filters \cite{zhang2023spatially} for deformable image registration. 
TextSCF harnesses the latent relationships among anatomical regions by utilizing large-scale visual-language models \cite{radford2021learning}, enhancing its understanding of contextual anatomy.

\section{Method}
\begin{figure*}[t]
    \centering
    \includegraphics[width=1.8\columnwidth]{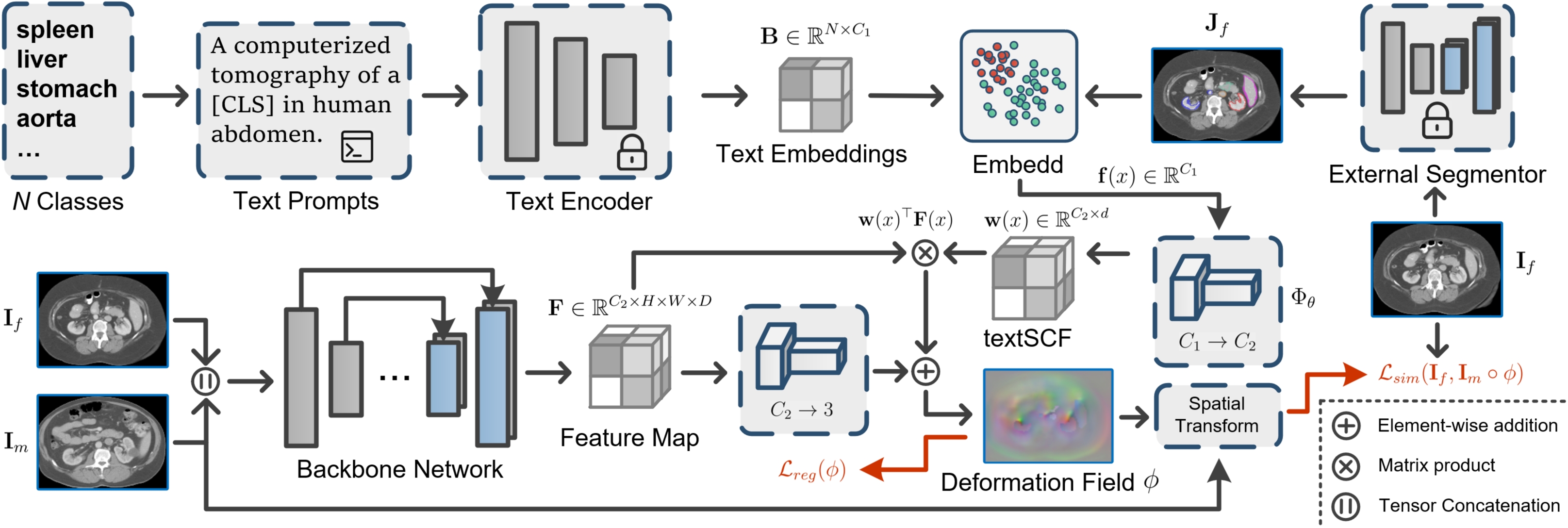}
    \caption{
        Visual illustration of the overall framework for the proposed textSCF is provided.
        The process details are described in Section~\ref{sec:framework}.
        Orange arrows indicate the loss functions (the details can be found in \ref{sec:loss_function}); note that the Dice loss is omitted in the figure for brevity. A lock symbol on a module means that its weights are frozen and not subject to training.
        } 
    \label{fig:textscf_framework}
\end{figure*}

\subsection{Preliminaries}
Deformable image registration determines voxel-level correspondences between a moving image $\mathbf{I}_m$ and a fixed image $\mathbf{I}_f$. 
The spatial mapping is denoted as $\mathbf{\phi}(x) = x + \mathbf{u}(x)$, with $x$ indicating a location in the domain $\Omega \subset \mathbf{R}^{H\times W\times D}$ and $\mathbf{u}(x)$ being the displacement vector at $x$.
The displacement field $\mathbf{u}(x)$ warps the moving image $\mathbf{I}_m$ to align with the fixed image $\mathbf{I}_f$, ensuring every voxel in $\mathbf{I}_f$ corresponds to a voxel in the warped $\mathbf{I}_m$ (denoted as $\mathbf{I}_m \circ \phi$), with trilinear interpolation calculating values for non-grid positions.
Unsupervised learning involves training a network $F_{\theta}$ to estimate the deformation field $\phi$ from $\mathbf{I}_m \circ \phi$ and $\mathbf{I}_f$: $\phi=F_{\theta}(\mathbf{I}_m,\mathbf{I}_f)$.
The network weights $\theta$ are optimized by minimizing a composite loss function $\mathcal{L}$, which combines measures of dissimilarity between warped $\mathbf{I}_m$ and $\mathbf{I}_f$, smoothness of the deformation field, and an auxiliary loss for enhanced alignment:
\begin{equation}
    \mathcal{L} = \mathcal{L}_{sim}(\mathbf{I}_f,\mathbf{I}_m \circ \phi) + \mathcal{L}_{auc}(\mathbf{J}_f,\mathbf{J}_m\circ \phi) + \lambda \mathcal{L}_{reg}(\phi),
    \label{eq:loss_general}
\end{equation}
where $\mathcal{L}_{sim}(\cdot,\cdot)$ quantifies the similarity, $\mathcal{L}_{reg}(\cdot)$ imposes regularization to ensure deformation smoothness, and $\mathcal{L}_{dsc}(\cdot,\cdot)$ serves as an auxiliary loss.
The coefficient $\lambda$ modulates the smoothness of the deformation field.
In our implementation, we employ the Mean Square Error (MSE) loss to gauge dissimilarity, while the smoothness of the displacement field is encouraged using the L2 norm of its spatial gradients $ ||\nabla \mathbf{u}||^2 $. 
Consistent with \cite{balakrishnan2019voxelmorph}, Dice loss evaluates alignment between the moving segmentation $\mathbf{J}_m \circ \phi$ and the fixed segmentation $\mathbf{J}_f$.

\subsection{Weakly-supervised Registration}
In fully-supervised image registration, training involves both input images and their matching ground-truth deformation fields, while inference is conducted with just the input images.
On the other hand, unsupervised image registration uses only input images for training and inference.
Our weakly-supervised method uses input images and segmentation masks during both training and inference, offering a middle ground between fully-supervised and unsupervised approaches.
Moreover, the segmentation masks employed during training can be sourced from either ground-truth data or generated automatically. 
For inference, the model relies solely on masks produced by external segmentation models, circumventing the need for additional efforts, aligning the process with the unsupervised methods.
In this study, we aim to train a network $F_{\theta}$ that processes the input images $\mathbf{I}_m$, and $\mathbf{I}_f$, along with the fixed segmentation mask $\mathbf{J}_f$ to predict the deformation field: $\phi=F_{\theta}(\mathbf{I}_m,\mathbf{I}_f,\mathbf{J}_f)$.

\subsection{The Overall Framework of textSCF}
\label{sec:framework}
The proposed text-driven registration framework consists of three core components: a feature branch with a standard encoder-decoder network, a mask branch utilizing an external segmentation model to locate anatomical regions, and a text branch that encodes text prompts associated with anatomical regions.
Next, we will detail each component within the framework.

\subsubsection{Text Branch}
Considering the impressive capabilities of few-shot learning \cite{radford2021learning} and modeling anatomical relationships \cite{liu2023clip} of pretrained visual-language models like CLIP \cite{brown2020language}, we aim to harness this knowledge by using CLIP to encode anatomical regions.
For the $k_{th}$ anatomical region, we generate a text prompt like ``\texttt{a photo of a [CLS].}'', where \texttt{[CLS]} is replaced with the region's name. 
We then employ CLIP to obtain the corresponding embedding vector $\mathbf{b}_k$ for that region.
If we have $N$ anatomical regions, the resulting vectors compose an embedding matrix $\mathbf{B} \in \mathbb{R}^{N\times C_1}$, where $C_1$ is length of each vector.

We carefully tailor the background vector to hedge against uncertainties in anatomical segmentation masks from external models. 
This vector is designed to have a resilient encoding that adjusts to regional variations and stands out against each distinct anatomical region.
Let $\tilde{\mathbf{B}}$ represent the matrix composed of last $N-1$ vectors.
We perform singular value decomposition (SVD) on $\tilde{\mathbf{B}}$, yielding $\tilde{\mathbf{B}} = \mathbf{U}\mathbf{\Sigma}\mathbf{V}^{\top}$.
To determine the background vector $\mathbf{b}_0$, we maximize its orthogonality to the subspace spanned by the rows of $\tilde{\mathbf{B}}$. 
Mathematically, $\mathbf{b}_0$ is the last column of $\mathbf{V}$, denoted as $\mathbf{b}_0=\mathbf{V}[:,-1]$.

\subsubsection{Mask Branch and the Derivation of textSCF}
Conventional registration networks apply identical filters across all pixel locations when generating the deformation field, which may not yield the best results. 
Displacement patterns often vary between regions, reflecting each subject's distinctive traits like organ placement and size. 
This results in displacement fields that are internally consistent within a region but can vary greatly between different regions.
Therefore, we introduce the concept of spatially covariant filters to accommodate these variations in regional displacement.

The original SCP approach \cite{zhang2023spatially} employs a neural network to establish an implicit function \cite{NeRF} that maps pixel coordinates to corresponding classifier weights:
\begin{equation}
\Phi_{\theta}: (x \in \mathbb{R}^d) \mapsto (\mathbf{w}(x) \in \mathbb{R}^C),
\label{eq:scp}
\end{equation}
where $x$ represents the pixel's coordinates in a $d$-dimensional space ($d=3$ in our volumetric registration), $\mathbf{w}(x)$ is the weight vector with $C$ elements, and $\theta$ is the trainable parameter (Note that $\mathbf{w}(x)$ can be a matrix of $C\times N$ if it is an $N$-label segmentation).
The method is effective for structured medical image segmentation but less so for registration due to the need for a dense deformation field in non-aligned data.
In this work, we suggest replacing pixel coordinates in Eq.~\eqref{eq:scp} with encoded vectors representing anatomical regions.

First, we apply a pretrained external segmentation model like SwinUnetr \cite{tang2022self} to the fixed image $\mathbf{I}_f$ to produce a probability distribution vector $\mathbf{d}(x) \in \mathbb{R}^{N}$ representing the $N$ anatomical regions for each location.
Then, we use both argmax and max operations on $\mathbf{d}(x)$ to derive the segmentation mask $\mathbf{J}_f(x)$ and extract the confidence score, represented by the probability $p(x)$.
Then, we apply $\mathbf{J}_f(x)$ to look up on $\mathbf{B}$ and obtain an embedding vector $\mathbf{f}(x)$.
Finally, we can derive the proposed textSCF as follows:
\begin{align}
    \mathbf{w}(x) &= \Phi_{\theta}(\mathbf{f}(x)), \\
    \mathbf{u}(x) &= (p(x)\cdot\mathbf{w}(x)+(1-p(x))\cdot\mathbf{w}_r)^{\top}\mathbf{F}(x), 
    \label{eq:textscf}
\end{align}
where $\mathbf{u}(x)$ is the displacement vector, $\mathbf{F}(x)\in \mathbb{R}^{C_2}$ is the feature vector produced by the backbone network, $\mathbf{w}(x) \in \mathbb{R}^{C_2\times d}$ is the estimated filter weights from implicit function $\Phi$, with $\theta$ as its trainable parameters.
$\mathbf{w}_r \in \mathbb{R}^{C_2\times d}$ is a trainable parameter (implemented with a linear layer) that acts as a uniform filter applied universally across all spatial locations.
Substituting $\mathbf{w}(x)$ with $\mathbf{w}_r$ in Eq. \eqref{eq:textscf} simplifies the network to perform translation-invariant filtering.
With the displacement vector, we can obtain the deformation field $\phi(x)=x+\mathbf{u}(x)$.
The implicit function $ \Phi_{\theta} $ is implemented using a multi-layer perceptron (MLP) comprising three layers. 
Each layer includes a linear layer coupled with a ReLU activation function, except for the final layer, which consists solely of a linear component.
Illustrated in Fig. \ref{fig:encoder_decoder}, $ \Phi_{\theta} $ transitions text embeddings from a $C_1$-dimensional space to a $C_2$-dimensional space.

\subsubsection{Feature Branch and Overall Framework}

\begin{figure}[t]
    \centering
    \includegraphics[width=1.0\columnwidth]{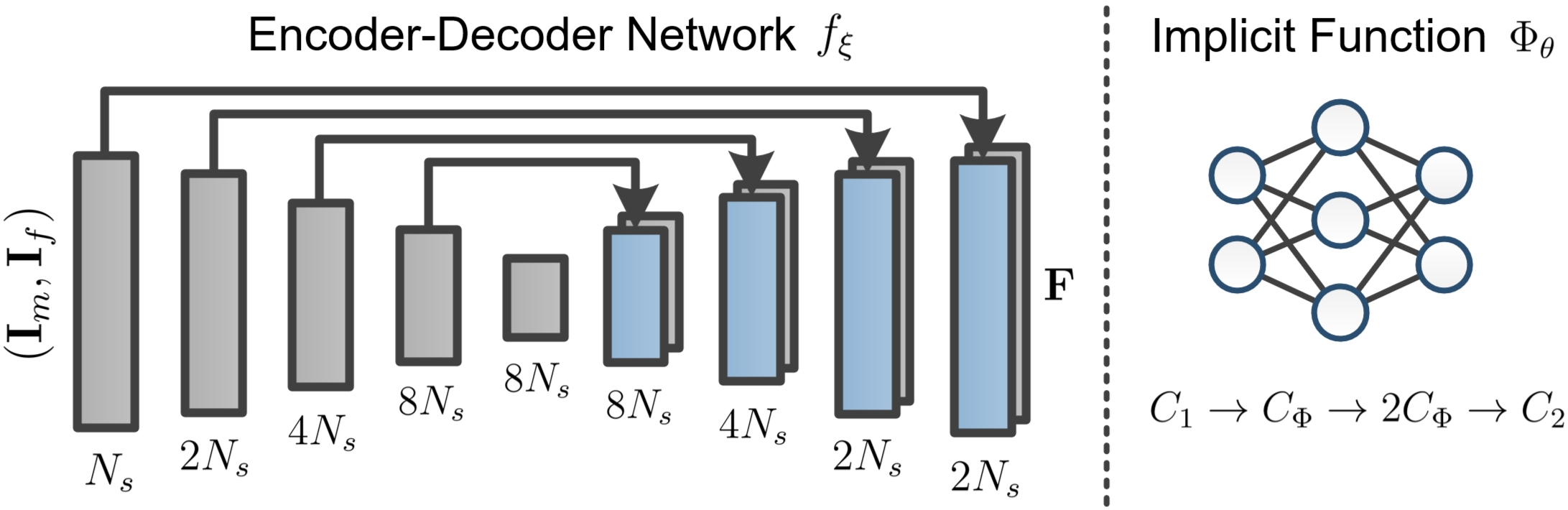}
    \caption{
        The figure demonstrates the structure of the encoder-decoder backbone network $f_{\xi}$ (left panel) and the architecture of the implicit function $\Phi_{\theta}$ (right panel). Feature tensors are represented by rectangles, each produced from its predecessor through a 3D convolutional layer. 
        Arrows signify skip connections, merging tensors of the encoder and decoder, where $N_s$ indicates the channel count. 
        The $\Phi_{\theta}$ function, realized by an MLP with three layers, maps text embeddings from dimension $C_1$ to successive intermediate dimensions $C_{\Phi}$ and $2C_{\Phi}$, culminating in the final dimension $C_2$.
    }
    \label{fig:encoder_decoder}
\end{figure}

The feature branch features a backbone network that adopts an encoder-decoder structure, similar to architectures like VoxelMorph \cite{balakrishnan2019voxelmorph} and its variants (See Fig. \ref{fig:encoder_decoder}). 
We represent this network as $f_{\xi}$, with $\xi$ denoting the network's trainable parameters.
This branch combines the moving image $\mathbf{I}_m$ and the fixed image $\mathbf{I}_f$ along the channel dimension to produce a feature map $\mathbf{F}\in \mathbb{R}^{C_2\times H \times W \times D}$ with $C_2$ channels and a spatial size of $H \times W \times D$, calculated by $\mathbf{F}=f_{\xi}(\mathbf{I}_m,\mathbf{I}_f)$.
Here $N_s$ specifies the starting number of channels, a parameter that will inform the construction of networks with varying complexities.


The overall framework of our textSCF model is depicted in Fig. \ref{fig:textscf_framework}. 
The process begins with generating text prompts from anatomical region names, such as ``\texttt{liver}'' and ``\texttt{spleen}'', using a specific template.
Our ablation study, outlined in Table \ref{tab:text_embedding}, shows that textSCF, with its semantic encoding, considerably outperforms the original SCP model, which uses only spatial encoding. 
This study underscores the critical role for the choice of text prompts. 
Although further investigation into optimal prompting strategies is a compelling direction for future research, it falls outside the scope of this paper.
Then these prompts are subsequently fed into a pretrained text encoder coupled with SVD to produce the text embeddings $ \mathbf{B} $.
A segmentation model inputs the fixed image $ \mathbf{I}_f $ to output the segmentation map $ \mathbf{J}_f $ and its probability distribution, which, combined with the text embeddings, are assigned to each voxel.
These combined embeddings $ \mathbf{f} \in \mathbb{R}^{C_1 \times H \times W \times D} $ are passed through a function $ \Phi_{\theta} $ to produce spatially covariant filters $\mathbf{w} \in \mathbb{R}^{C_2 \times d \times H \times W \times D} $.
Finally, with the feature map from $ f_{\xi} $, the deformation field is calculated as outlined in Eq.~\eqref{eq:textscf}.

\subsection{Loss Function}
\label{sec:loss_function}
Unless otherwise specified, all methods apart from LapIRN utilize the same loss function with $\lambda=0.1$, mirroring the textSCF framework. 
For LapIRN, to attain peak performance, we employed the Normalized Cross-Correlation (NCC) loss for dissimilarity measurement and adjusted $\lambda$ to 3.5, in line with what is mentioned in its original code repository \footnote{\url{https://github.com/cwmok/LapIRN}}. 
Additionally, we integrated a Dice loss ${L}_{dsc}$ into the LapIRN setup. Therefore, the final loss function $\mathcal{L}$ is formulated as,
\begin{equation}
    \mathcal{L} = \mathcal{L}_{mse}(\mathbf{I}_f,\mathbf{I}_m \circ \phi) + \frac{1}{n}\sum \mathcal{L}_{dsc}(\mathbf{J}_{fi},\mathbf{J}_{mi}\circ \phi) + \lambda  ||\nabla \mathbf{u}||^2,
    \label{eq:loss_final}
\end{equation}
where the $\mathcal{L}_{mse}(\cdot,\cdot)$ denotes the MSE loss between the warped moving image and fixed image, $i$ denotes the $i$-th segmentation label and $n$ is the number of segmentation labels.

\begin{table}[t!]
\centering
\resizebox{1.\columnwidth}{!}{
\begin{tabular}{l|c|c|c} 
\hline
\hline
\textbf{Text Prompt} & \textbf{Embedding} & \textbf{Backbone} & \textbf{DSC(\%)} \\ \hline
\# 0: \texttt{-} & SCP \cite{zhang2023spatially} & - & 51.56\\
\# 1: \texttt{-} & One-Hot & - & 59.05 \\
\# 2: \texttt{A photo of a [CLS].} & CLIP  &  ViT & 59.33 \\
\# 3: \texttt{A photo of a [CLS].} & CLIP + SVD  & ViT & 60.17 \\
\# 4: \texttt{A photo of a [CLS] in human abdomen.} & CLIP + SVD  & ViT & 60.61  \\
\# 5: \texttt{A computerized tomography of a [CLS]} & \multirow{2}{*}{CLIP + SVD}  & \multirow{2}{*}{ResNet} & \multirow{2}{*}{60.01} \\
~~~~~~ \texttt{in human abdomen.} & &  \\
\# 6: \texttt{A computerized tomography of a [CLS]} & \multirow{2}{*}{CLIP + SVD}  & \multirow{2}{*}{ViT} & \multirow{2}{*}{\textbf{60.75}} \\
~~~~~~ \texttt{in human abdomen.} & &  \\
\# 7: \texttt{A computerized tomography of a [CLS]} & \multirow{2}{*}{ChatGPT + SVD}  & \multirow{2}{*}{-} & \multirow{2}{*}{57.63} \\
~~~~~~ \texttt{in human abdomen.} & &  \\
\hline
\end{tabular}
}
\vspace{1ex}
\caption{
    This ablation study examines the impact of textual anatomical embeddings on performance. ``CLIP+SVD" refers to the use of CLIP for encoding anatomical text prompts, with SVD for background vector encoding. ``ViT" utilizes the `ViT-L/14@336px' model from the CLIP family as a representative of large transformer architectures, whereas ``ResNet" uses the `RN50x64' model to represent large convolutional networks. For ``ChatGPT", we employ the `text-embedding-ada-002' model for text embeddings. The reported metric is the average Dice on the Abdomen dataset.
}
\label{tab:text_embedding}
\end{table}

\section{Experiments and Results}
\subsection{Datasets}
\subsubsection{OASIS}
We conducted inter-subject brain registration using the Learn2Reg registration challenge 2021 \cite{hering2022learn2reg}, which utilizes the OASIS dataset \cite{marcus2007open}.
This dataset comprises T1w brain MRI scans from 414 subjects. 
For training, we used 394 unpaired scans, and 19 image pairs from 20 scans were employed for validation and public leaderboard ranking \footnote{\url{https://learn2reg.grand-challenge.org/evaluation/task-3-validation/leaderboard/}}.
In our experiments, we utilized pre-processed data from the challenge organizers, which included bias correction, skull stripping, alignment, and cropping of all scans to a size of $160\times192\times224$.

\subsubsection{Abdomen CT}
We also performed inter-subject organ registration using the Abdomen CT dataset~\cite{xu2016evaluation} from the Learn2Reg~\cite{hering2022learn2reg} challenge 2020.
This dataset includes 30 abdominal CT scans, with each scan segmented into 13 anatomical structures.
All images were resampled to a consistent voxel resolution of 2 mm and a spatial size of $192\times160\times256$.
We divided the dataset into three parts: 20 CT images for training, 3 for validation, and 7 for testing, which results in 380 ($20\times19$) training pairs, 6 ($3\times2$) validation pairs, and 42 ($7\times6$) testing pairs.

\subsubsection{Why Choose Them?}
The selected datasets present anatomical regions with various shapes, sizes, and locations, making them ideal for evaluating our method. 
They contain two commonly used imaging modalities: MRI and CT, each with its own set of challenges. 
The main difficulty of the OASIS dataset lies in fine-grained alignments of small and variably shaped brain structures. 
The Abdomen CT dataset, on the other hand, is primarily challenging due to large deformations and its relatively small size. 
Assessing on these datasets offers a more comprehensive and convincing evaluation.

\subsection{Implementation Details and Baseline Methods}
\subsubsection{Training Details}
All models were developed using PyTorch in Python \cite{paszke2019pytorch}. 
The training environment included a machine with 32GB memory, a 16-core CPU, and an A100 GPU. 
Network training utilized the Adam optimizer \cite{kingma2014adam} with an initial learning rate of $1e-4$, complemented by a polynomial learning rate scheduler with a 0.9 decay rate.
The training process was set to a batch size of 1 and continued for 700 epochs for the OASIS dataset and 100 epochs for the Abdomen CT dataset. 
Throughout the paper, we set $\lambda=0.1$ for the smoothness regularization term, except where noted otherwise.
For a fair comparison, all models were trained either under same conditions or according to their specified preferred settings in their repositories.

\subsubsection{Data Processing}
Our approach involved a straightforward data processing pipeline. 
In line with the learn2reg challenge guidelines \cite{hering2022learn2reg}, the output deformation field is spatially halved for both datasets. 
For the OASIS dataset, we maintained the original size of the input images, whereas, for CT datasets, both the input and output sizes are halved. 
We normalized all images to ensure their intensities fall within the range of $[0,1]$. 
Specifically for the CT dataset, intensities were clipped between $-500$ and $800$ Hounsfield units prior to the normalization.
For the OASIS dataset, we utilized automated segmentation masks generated by FreeSurfer \cite{fischl2012freesurfer} and the Neurite package \cite{dalca2018anatomical} for both the Dice loss computation and as inputs to textSCF. 
In the Abdomen CT dataset, manual segmentation masks were employed for calculating the Dice loss, while automated segmentation masks obtained from the pretrained SwinUnetr model \cite{tang2022self} served as inputs to textSCF.

\subsubsection{Baseline Methods and Model Details of textSCF}
Our study compares the textSCF with several state-of-the-art non-iterative, learning-based baseline models, including VoxelMorph \cite{balakrishnan2019voxelmorph}, LapRIN \cite{mok2020large}, TransMorph \cite{chen2022transmorph}, LKU-Net \cite{jia2022u}, and Fourier-Net \cite{jia2023fourier}. 
For the OASIS dataset, we obtained evaluation scores from the public leaderboard or respective publications. 
In the case of the Abdomen CT dataset, we acquired the models' code from their public repositories and fine-tuned each to achieve optimal performance.
Although textSCF is model-agnostic, we opted for the LKU-Net backbone due to its simplicity and effectiveness in capturing both fine-grained details and large deformations. 
For all our experiments, we standardized the size of the kernel to 5. 

\subsubsection{Evaluation Metrics}
Consistent with established methods \cite{dalca2018unsupervised,balakrishnan2019voxelmorph,mok2020large,chen2022transmorph} and challenge protocols \cite{hering2022learn2reg}, our evaluation metrics include the Dice Similarity Coefficient (Dice) and the $95 \%$ percentile of the Hausdorff Distance (HD95) for similarity assessment of anatomical regions.
To evaluate the diffeomorphism quality of deformation fields, we used the standard deviation of the logarithm of the Jacobian determinant (SDlogJ). 
Furthermore, to assess computational complexity, we measured the network's parameter size and multi-add operation count for each method.



\begin{table}[!t]

\begin{center}
\resizebox{0.9\columnwidth}{!}{
\begin{tabular}{ lccc }
\hline
\hline
Model &  Dice (\%) $\uparrow$ & HD95 (mm) $\downarrow$ & SDlogJ $\downarrow$ \\ 
\hline
Initial & 57.18 & 3.83 & - \\
VoxelMorph \cite{balakrishnan2018unsupervised} & 84.70 & 1.55 & 0.13 \\
Fourier-Net \cite{jia2023fourier} & 86.04 & 1.37 & 0.48 \\
LapIRN \cite{mok2020large} & 86.10 & 1.51 & \textbf{0.07} \\
TransMorph \cite{chen2022transmorph} & 86.20 & 1.43 & 0.13 \\
TM-TVF \cite{chen2022unsupervised} & 87.06 & 1.39 & 0.10 \\
LKU-Net \cite{jia2022u} & 88.61 & 1.26 & 0.52 \\
im2grid \cite{liu2022coordinate} & 89.35 & 1.24 & 0.23 \\
textSCF (ours) & \bf{90.05} & \bf{1.22} & 0.39 \\
\hline
\end{tabular}
}
\end{center}
\caption{
Quantitative comparison of registration results on the OASIS dataset, featuring our textSCF method alongside other techniques. 
Metrics including Dice (\%), HD95 (mm), and SDlogJ are averaged across all image pairs for each method. 
The best-performing metrics in each category are denoted in bold. 
Symbols indicate the desired direction of metric values: $\uparrow$ implies higher is better, while $\downarrow$ indicates lower is better. 
``Initial'' refers to the baseline results before registration.
}
\label{tab:brain_reg}
\end{table}

\subsection{Results and Analysis}

\subsubsection{Analysis of Textual Anatomical Embeddings}
Our ablation study (Table \ref{tab:text_embedding}) underscores text encoding's importance in textSCF's registration accuracy. 
Location-specific terms like ``in human abdomen" enhance Dice scores by 0.7\%, while specifying the imaging modality like ``a computerized tomography" provides further improvement. 
The ViT model as a backbone surpasses ResNet, particularly evident in prompts ``\#5" and ``\#6". 
SVD's role in background encoding, as in prompts ``\#2" and ``\#3", proves beneficial. 
However, ChatGPT embeddings (prompt ``\#7") fall short compared to CLIP's image-text pairings, and SCP's sole spatial encoding is less effective. 
The study validates textSCF's strength in combining SCFs with text prompts for precise anatomical region-specific filtering and capturing semantic relationships between regions.

\begin{table}[!t]
\begin{center}
\resizebox{0.9\columnwidth}{!}{
\begin{tabular}{ lccc }
\hline
\hline  
Model &  Dice (\%) $\uparrow$ & HD95 (mm) $\downarrow$ & SDlogJ $\downarrow$ \\ 
\hline
Initial & 30.86 & 29.77 & - \\
VoxelMorph \cite{balakrishnan2018unsupervised} & 38.64 & 25.58 & 0.24 \\
TransMorph \cite{chen2022transmorph} & 39.40 & 23.71 & 0.29 \\
Fourier-Net \cite{jia2023fourier} & 40.55 & 23.70 & \bf{0.16} \\
LKU-Net \cite{jia2022u} & 50.32 & \bf{20.20} & 0.35 \\
LapIRN \cite{mok2020large} & 54.55 & 20.52 & 1.73 \\
textSCF (ours) & \bf{60.75} & 22.44 & 0.87 \\
\hline
\end{tabular}
}
\end{center}
\caption{
Quantitative comparison on the Abdomen CT dataset, comparing textSCF with other methods. 
Metrics Dice (\%), HD95 (mm), and SDlogJ are averaged for each method. 
Top results in each metric are in bold. 
}
\label{tab:abdomen_reg}
\end{table}

\begin{figure*}[!ht]
    \centering
    \includegraphics[width=2.1\columnwidth]{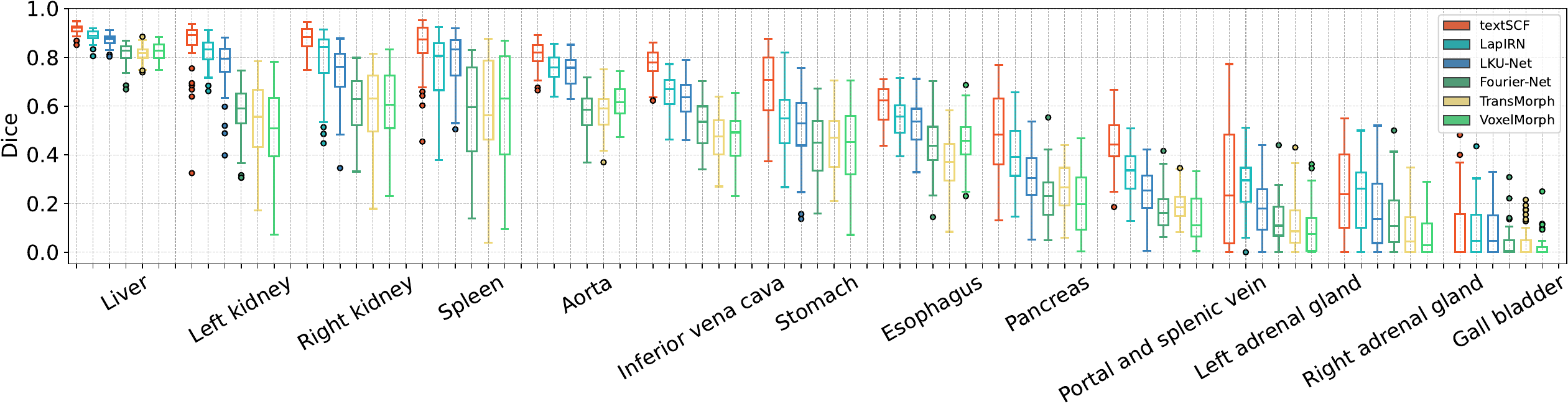}
    \caption{
        Boxplots depicting the Dice scores for each anatomical structure on the Abdomen CT dataset. 
        Included structures are the spleen, right kidney, left kidney, gall bladder, esophagus, liver, stomach, aorta, inferior vena cava, portal and splenic vein, pancreas, left adrenal gland, and right adrenal gland. 
        Structures are arranged in order of their average Dice score achieved with textSCF.
    }
    \label{fig:abdomen_boxplot}
\end{figure*}

\subsubsection{Registration Accuracy}

\begin{figure}[!ht]
    \centering
	\subfloat[Brain]{\includegraphics[width=.49\columnwidth]{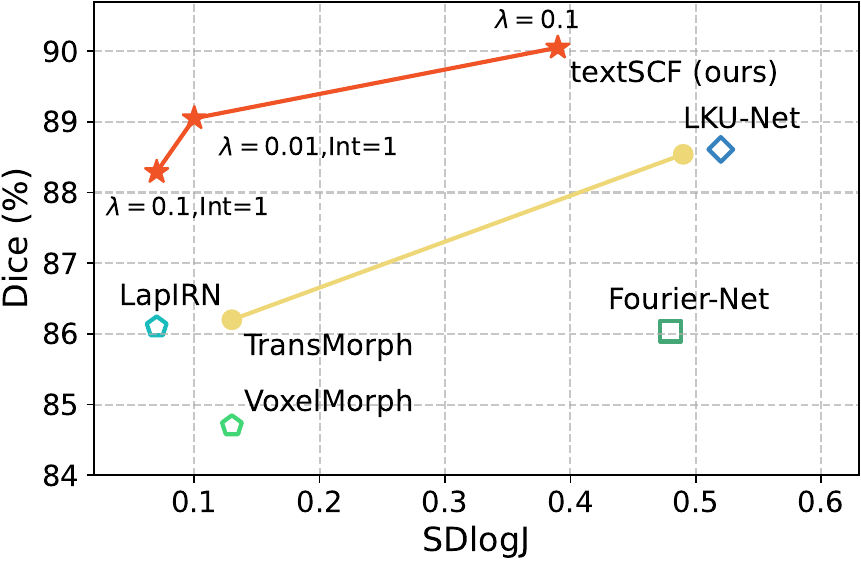} \label{fig:oasis_smoothness}} 
    \subfloat[Abdomen]{\includegraphics[width=.49\columnwidth]{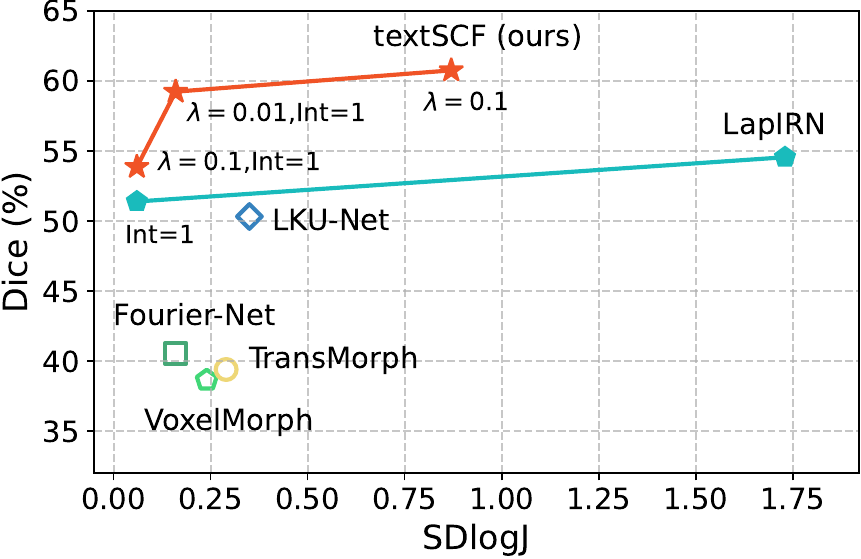}\label{fig:abdomen_smoothness}}
    \caption{
        Trade-off between smoothness and Dice (\%).
        This plot shows the relationship between average Dice scores and the smoothness metric SDlogJ in brain and abdomen registrations. 
        In both registrations, network variants vary by the application of a diffeomorphic integration layer \cite{dalca2018unsupervised} (indicated by ``Int'' next to points) and adjustments in the global smoothness term coefficient $\lambda$.
    }
    \label{fig:smoothness}
\end{figure}

Table \ref{tab:brain_reg} presents the quantitative results of our proposed textSCF method in comparison to other methods on brain registration with the OASIS dataset. 
In summary, textSCF demonstrates superior performance in both the average Dice score and HD95 score while maintaining comparable smoothness in the deformation field (see Fig. \ref{fig:oasis_smoothness} for accuracy-smoothness trade-off). 
Notably, our model, which incorporates a text encoder and spatially covariant filters, shows improvement in all three metrics over the LKU-net, its direct counterpart without these features.

In Table \ref{tab:abdomen_reg}, we present quantitative results comparing our textSCF method to others in inter-subject abdomen registration. 
Among methods that exhibit smoother deformation fields than textSCF (indicated by a lower SDlogJ), textSCF stands out with significantly better Dice scores.
Specifically, textSCF achieves improvements in Dice score of 57.17\% over VoxelMorph, 54.17\% over TransMorph, 49.94\% over Fourier-Net, and 20.72\% over LKU-Net. 
Although LapIRN \cite{mok2021large} is optimized for large deformations and leads in Dice score, its higher SDlogJ suggests less smoothness in deformation. 
Notably, textSCF shows an 11.37\% improvement in Dice score compared to LapIRN, while also maintaining a smoother deformation field.

To gain deeper insights into the performance of various deformable registration methods,  we visualized Dice score distributions across anatomical structures in the Abdomen CT dataset, as seen in Fig. \ref{fig:abdomen_boxplot}. 
The proposed textSCF consistently outperforms other methods across these anatomical structures. 
Notably, in a paired t-test, textSCF significantly surpasses Fourier-Net, TransMorph, and VoxelMorph in Dice scores for all structures, with statistical significance ($p<0.05$). 
However, no statistical significance was found for the gall bladder in comparisons with LKU-Net and LapIRN, and for the left and right adrenal glands with LapIRN.
Moreover, all methods, including textSCF, show reduced performance on the left/right adrenal gland and gall bladder, attributed to their small size and irregular shapes.

\subsection{Smoothness Analysis}

In deformable image registration, a diffeomorphism enables smooth, complete image transformations without tearing or folding, preserving topologies. 
Smoothness is typically encouraged in learning frameworks via regularizers or a diffeomorphic integration layer \cite{dalca2018unsupervised}. 
However, prioritizing smoothness may impact anatomical correspondence accuracy, making it crucial to find a balance between the two.

The Fig. \ref{fig:smoothness} displays two scatter plots comparing different registration methods in terms of their Dice scores and smoothness measure SDlogJ for both brain and abdomen registrations.
It highlights that textSCF leads in Dice scores, demonstrating its unmatched registration accuracy among the compared methods. 
Even with comparable levels of smoothness, textSCF maintains higher Dice scores, illustrating its proficiency in finding anatomical correspondences. 
LKU-Net and LapIRN, while individually strong in either smoothness or accuracy, lack this dual advantage. 
Consequently, textSCF emerges as the method with the best balance between registration accuracy and smoothness.

\begin{figure}[t]
    \centering
    \subfloat[Impact of $C_{\Phi}$ on Accuracy]{\includegraphics[width=.5\columnwidth]{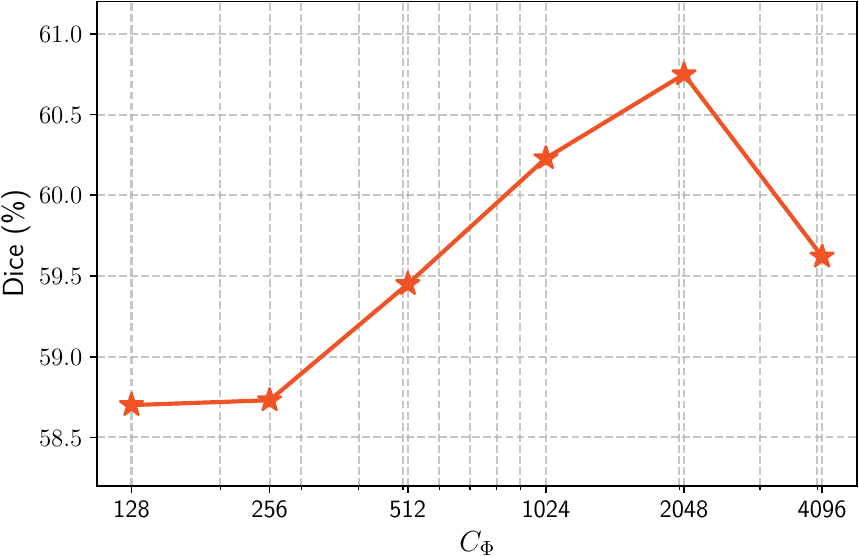}\label{fig:phi_channels}}
    \subfloat[Accuracy vs. Complexity]{\includegraphics[width=.5\columnwidth]{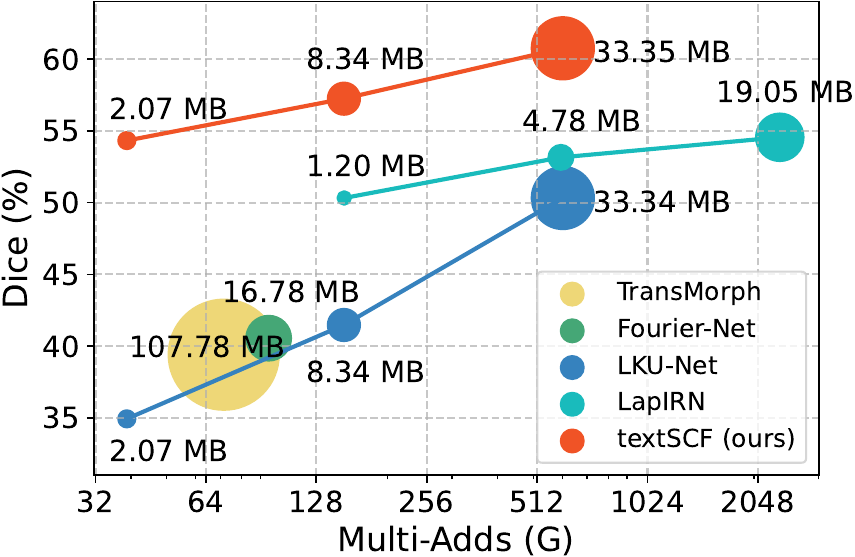} \label{fig:abdomen_complexity}} 
    \caption{
        (a) This graph depicts the correlation between the channel count \( C_{\Phi} \) and the Dice score achieved by textSCF on the Abdomen dataset. 
        Each data point corresponds to the Dice score attained at varying levels of \( C_{\Phi} \), which is plotted on a logarithmic scale to highlight the incremental enhancements.
        (b) This plot illustrates the trade-off between average Dice (\%) and computational complexity for the abdomen dataset. 
        It compares network parameter size and multi-add operations (in G), with the x-axis on a logarithmic scale. 
        Starting channel counts \( N_s \) for textSCF, LapIRN, and LKU-Net increase from 8 to 32, correlating with left-to-right movement on the graph. 
        Circle size indicates the parameter size of each network.
    }
    \label{fig:phi_channels_complexity}
\end{figure}


\subsection{Complexity Analysis}
The complexity of the textSCF model is governed by two parameters: the starting number of channels, $N_s$, in the backbone network $f_{\xi}$, and the number of channels, $C_{\Phi}$, in function $\Phi_{\theta}$.
As network complexity increases, so typically does registration accuracy, albeit at the cost of increased computational demands, such as larger network size and more multi-add operations.
Hence, a well-designed registration method need to have a careful balance between these aspects. 

\begin{table}[!t]

\begin{center}
\resizebox{1.0\columnwidth}{!}{
\begin{tabular}{ ccccc }
\hline
\hline
Lung Data & ``\texttt{Lung}'' Encode &  Dice (\%) $\uparrow$ & HD95 (mm) $\downarrow$ & SDlogJ $\downarrow$ \\ 
\hline
- & - & 60.75 & 22.44 & 0.87 \\
\checkmark & - & 59.81 & 22.17 & 0.79 \\
\checkmark & \checkmark & 62.07 & 22.50 & 0.86 \\

\hline
\end{tabular}
}
\end{center}
\caption{
    Assessing the effect of Lung CT data and ``Lung" encoding on registration metrics: Dice (\%), HD95 (mm), and SDlogJ. 
    Checkmarks indicate usage.
}
\label{tab:transferability}
\end{table}

\subsubsection{Effects of $C_{\Phi}$}
The impact of $C_{\Phi}$ on model performance was examined.
A higher $C_{\Phi}$ during the training phase may result in increased computational demands. 
However, during inference, $C_{\Phi}$ does not add to the complexity; once the model is trained, SCFs can be obtained in a single forward pass and stored for subsequent use with negligible additional cost. 
As illustrated in Fig. \ref{fig:phi_channels}, escalating $C_{\Phi}$, initially enhances the Dice score, with a peak performance at 2048 channels. 
Beyond this point, there is a marginal decline, suggesting that there is an optimal range of $C_{\Phi}$ for this model's architecture.

\subsubsection{Effects of $N_s$}
The complexity of our backbone network $f_{\xi}$ is modulated by the starting channel count $N_s$, as illustrated in Fig. \ref{fig:encoder_decoder}.
This parameter can also be used to govern the complexity of comparators such as LapIRN and LKU-Net.
Generally, a higher $N_s$ enhances registration accuracy.
Yet, for the OASIS dataset, performance declines when $N_s$ exceeds 64. 
Similarly, for the Abdomen dataset, a $N_s$ beyond 32 leads to lower accuracy.
This optimal $N_s$ threshold is linked to dataset size; an oversized network is prone to overfitting when data is limited.
Fig. \ref{fig:abdomen_complexity} presents an accuracy-complexity comparison of textSCF with other registration methods.
The proposed textSCF attains the most favorable balance: with comparable accuracy to LapIRN ($N_s$=32), textSCF ($N_s$=8) achieves an 89.13\% reduction in network parameter size and a 98.34\% reduction in multi-add operations.


\begin{figure}[t]
    \centering
    \includegraphics[width=0.8\columnwidth]{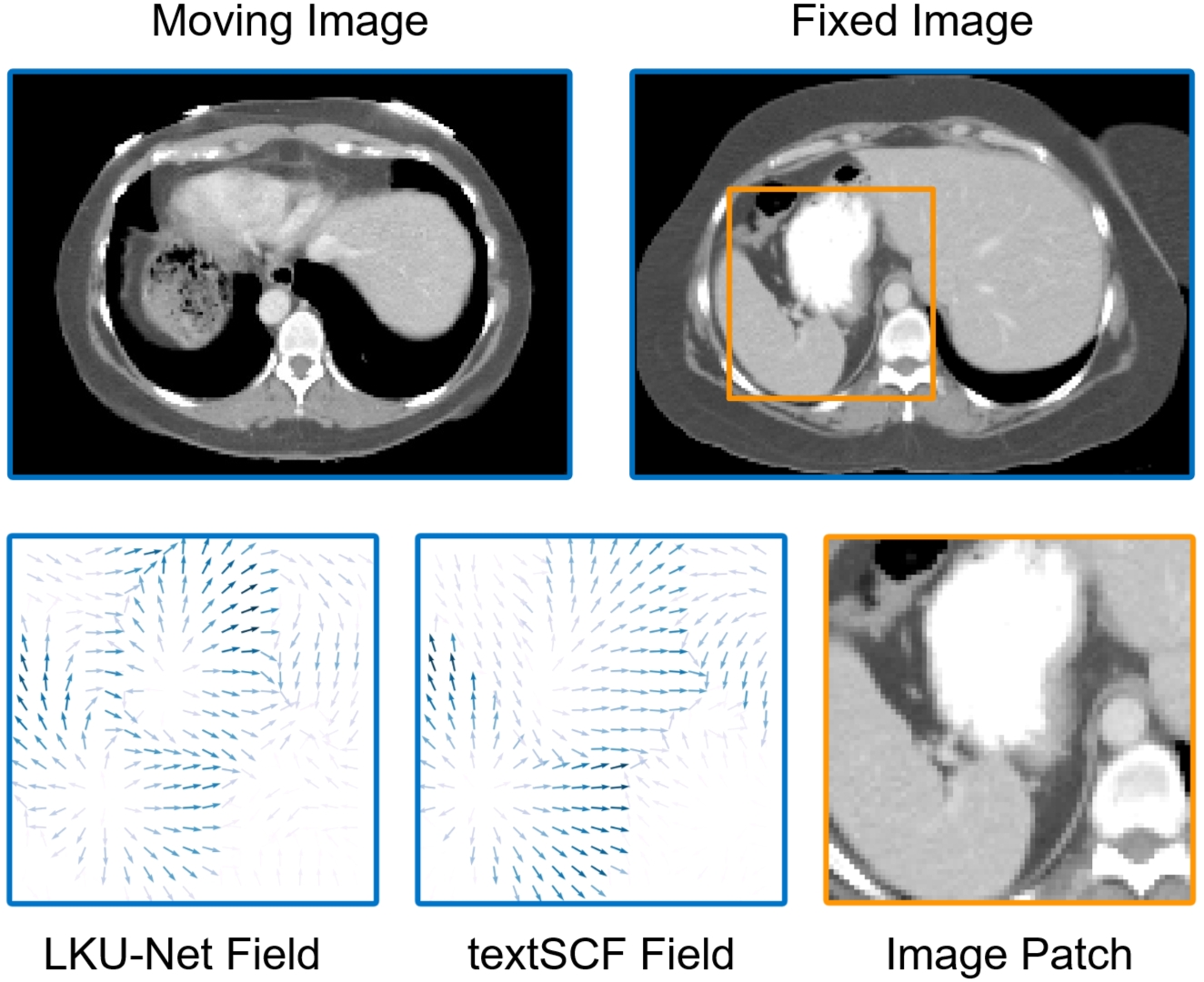}
    \caption{
        This figure contrasts deformation fields on an abdominal CT: LKU-Net's field (bottom left) blends boundaries, while textSCF's field (bottom middle) clearly marks the discontinuity, as seen in the highlighted area of the fixed image (top right and bottom right).
        The darker of the arrow, the larger of the displacement.
    }
    \label{fig:discontinuity}
\end{figure}

\subsection{Inter-Regional Transferability}
Our textSCF model demonstrates the capacity to transfer knowledge from external datasets with different anatomical regions, leveraging spatially covariant text embedding. 
We augmented abdomen CT registration with an auxiliary Lung CT dataset \cite{hering10learn2reg}. 
We processed 20 inspiration-phase lung images to align with the Abdomen CT dataset's specifications. 
During training, these lung images were combined with abdomen data, employing the same loss function. 
Table \ref{tab:transferability} reveals that adding Lung data without textual anatomical encoding marginally decreased performance. 
In contrast, specific text encoding of the lung as an additional anatomical region enhanced outcomes, highlighting the model's transferability capabilities through text encoding.
This indirectly showcases the model's ability to capture semantic relationships between different anatomical regions.

\subsection{Discontinuity-Preserving Capability}
Most current learning-based registration methods presuppose a globally smooth deformation field, which may not hold true for cases involving large deformations, such as abdominal registrations. Ideally, a deformation field should be smooth within each anatomical region but allow for discontinuities between different regions. 
This ability to preserve discontinuities is a crucial feature for a registration method. 
As depicted in Fig. \ref{fig:discontinuity}, textSCF exhibits this discontinuity-preserving capability, clearly delineating the stomach from surrounding areas, in contrast to LKU-Net's approach, which smooths over such boundaries.

\subsection{Expansibility Across Architectures}
\label{sec:univ_app}
The design of textSCF facilitates its integration with a variety of backbones. 
We tested its adaptability by implementing it with comparator baseline models as backbones, including VoxelMorph, TransMorph, LapIRN, and LKU-Net. 
These models were evaluated on the Abdomen dataset, notable for its large deformations. 
For each network's textSCF variant, we configured \( C_{\phi} \) to 2048 and employed the `base' version of the pretrained SwinUnetr \cite{tang2022self} as the external segmentor.


\begin{figure}[!t]
    \centering
    \includegraphics[width=0.9\columnwidth]{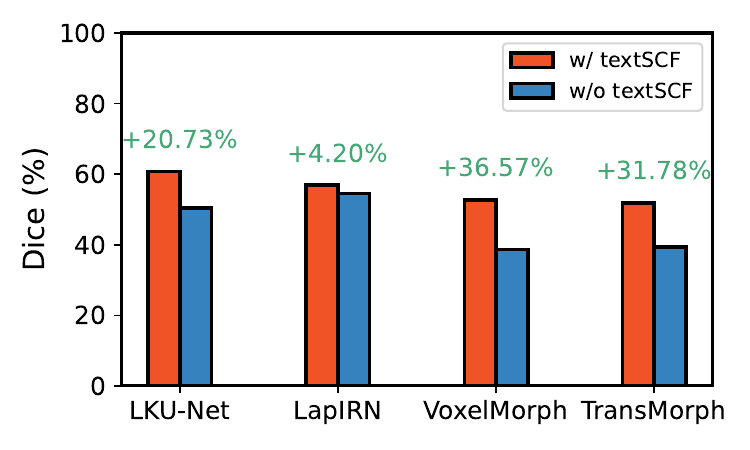}
    \caption{
        Comparative bar charts displaying the impact of integrating textSCF with various backbone networks. 
        The percentage values above the bars are the improvement achieved by incorporating textSCF into each model. 
    }
    \label{fig:dice_bar}
\end{figure}

Fig. \ref{fig:dice_bar} showcases the impact of the textSCF module on various backbone networks in terms of Dice (\%) metrics, highlighting textSCF's expansibility across architectures. 
The integration of textSCF leads to accuracy enhancements across all models, with VoxelMorph benefiting the most significantly. 
VoxelMorph outperforms TransMorph with textSCF integration, indicating a potentially greater synergy with ConvNet architectures, possibly due to ConvNets' implicit positional encoding via zero padding \cite{islam2019much}. 
This intrinsic characteristic of ConvNets may explain their superior performance over vision transformers in handling large deformations. 
Additionally, LapIRN's gains with textSCF are less marked than others.

\subsection{Impact of External Segmentation Accuracy}
\label{sec:effects_seg}
As previously noted, textSCF's registration accuracy is correlated with the accuracy of external segmentation inputs. 
We explore how changes in segmentation accuracy impact registration outcomes. 
Our evaluation on the Abdomen dataset involved using ground-truth masks with textSCF, modulating segmentation accuracy by adjusting the pretrained SwinUnetr network size from `base' to `small' and `tiny' variants.

\begin{table}[!t]

\begin{center}
\resizebox{1.0\columnwidth}{!}{
\begin{tabular}{ lc|ccc }
\hline
\hline
Segmentor & Accuracy (\%) &  Dice (\%) $\uparrow$ & HD95 (mm) $\downarrow$ & SDlogJ $\downarrow$ \\ 
\hline
Ground truth     & 100.00 & 74.73 & 20.13 & 0.86 \\
SwinUnetr base   & 77.92  & 60.75 & 22.44 & 0.87 \\
SwinUnetr small  & 74.96  & 58.53 & 21.36 & 0.79 \\
SwinUnetr tiny   & 63.35  & 56.88 & 20.09 & 0.72 \\
\multicolumn{1}{c}{-}               & -      & 50.32 & 20.20 & 0.35 \\
\hline
\end{tabular}
}
\end{center}
\caption{
    The table compares registration performance using different segmentation inputs with textSCF on the Abdomen dataset. 
    `Segmentor' refers to the source of segmentation masks, with `Accuracy' indicating the segmentation quality. 
    `Dice', `HD95', and `SDlogJ' are metrics used to assess the registration performance, where higher Dice scores and lower HD95 and SDlogJ values are desirable. 
}
\label{tab:external_acc}

\end{table}

We evaluated SwinUnetr's performance on the Abdomen dataset, calculating the average Dice score across all 30 instances as the metric for segmentation accuracy. 
Table \ref{tab:external_acc} illustrates the relationship between registration and segmentation accuracy. 
With ground-truth segmentation masks, the registration accuracy in terms of Dice (\%) reaches 74.73\%, significantly outperforming the SwinUnetr variants. 
As expected, registration accuracy decreases with segmentation accuracy for SwinUnetr variants. 
Intriguingly, as segmentation accuracy drops, both HD95 (mm) and SDlogJ metrics improve, likely because decreased segmentation precision leads to less pronounced structural discontinuities, resulting in a smoother deformation field.
Notably, textSCF enhances registration accuracy by 13.04\% over its non-textSCF variant, even when segmentation accuracy is as low as 63.35\%.

\subsection{Ablation Study}
To analyse the contribution of each component, we build several variations for textSCF, with the corresponding results in Table~\ref{tab:consolidated}. It can be found that, incorporating SCP~\cite{zhang2023spatially} on the baseline network leads to a 2.5\% improvement in registration Dice score. However, with our proposed SCF, the registration Dice rises to 59.05, significantly outperforming SCP. By further incorporating the text prompt, the registration Dice score of our textSCF rises to 60.75, demonstrating the efficiency of the text prompt. Our textSCF can also enhance inter-regional transferability, facilitating training on mixed source datasets (like human abdomen and lung CT) and yielding enhanced performance in target datasets (such as abdomen CT), evidenced by a Dice boost from 60.75 to 62.07.



\setcounter{table}{4}
\begin{table}[!h]
\begin{center}
\resizebox{1.0\columnwidth}{!}{
\begin{tabular}{ llc }
\hline
\hline
Location & Description &  Dice (\%) $\uparrow$ \\ 
\hline

Table 3, LKU-Net & Only baseline network & 50.32 \\
Table 1, row \#0 & Baseline network with only SCP [62] & 51.65 \\
Table 1, row \#1 & Baseline network with only SCF & 59.05 \\
Table 1, row \#6 & Baseline network with both text prompts and SCF & 60.75 \\
Table 4, row 3   & Baseline network with textSCF and extra lung CT datasets  & 62.07 \\

\hline
\end{tabular}
}
\end{center}
\caption{
Consolidated results for identifying improvement sources and the ablation study of textSCF using the abdomen dataset. 
}
\label{tab:consolidated}
\end{table}

\vspace{-2ex}
\section{Conclusion}
In developing textSCF, we noted two primary limitations.
Firstly, while many open-source and pretrained segmentation models are available, the registration accuracy is somewhat reliant on the precision of these external segmentations, with performance declining in tandem.
Secondly, textSCF's effectiveness is less pronounced in datasets with simpler structures, demonstrated by only a slight Dice increase in cardiac registration compared to its counterpart without textSCF.

In this paper, we introduced textSCF, a novel method for deformable medical image registration, addressing two key questions from Section \ref{sec:intro}. 
For \textbf{Q1}, we utilized internal prior information with anatomical-region specific filters to enhance intra-region consistency and inter-region distinction. 
For \textbf{Q2}, we harnessed external knowledge via pretrained segmentation models and CLIP, capturing semantic inter-regional relationships. 
textSCF showed remarkable performance in brain MRI and abdominal CT registration tasks, achieving top ranks in MICCAI Learn2Reg 2021 challenge and notable improvements.




\bibliographystyle{ieeetr}
\bibliography{references}

\newpage

\noindent\textbf{\large Appendix Summary}

\vspace{2ex}
\noindent In this supplementary material, we offer further insights into the textSCF model as well as other baseline models. 
Appendix \ref{sec:impl_details} offers expanded descriptions and implementation specifics of these comparative models. 
Lastly, Appendix \ref{sec:diff_reg} elaborates on diffeomorphic registration, covering definitions of diffeomorphisms, differentiable diffeomorphic integration layers, and the diffeomorphism quality metric SDlogJ.

\section{Implementation details}
\label{sec:impl_details}
We compared textSCF with multiple baseline models, and in this section, we provide more descriptions about the model and more details of how we implement them.

\vspace{1ex}
\noindent\textbf{VoxelMorph \cite{balakrishnan2019voxelmorph}.}
VoxelMorph stands at the forefront of unsupervised learning in medical image registration via ConvNets. 
As detailed in \cite{balakrishnan2019voxelmorph}, it comes in two variants: VoxelMorph-1 and the more advanced VoxelMorph-2, which doubles the feature counts of its predecessor. 
Our focus was on VoxelMorph-2, which, by expanding network size, achieves better performance compared to VoxelMorph-1.

\vspace{1ex}
\noindent\textbf{TransMorph \cite{chen2022transmorph}.}
TransMorph, built on the Swin transformer framework \cite{liu2021swin}, is known for its high registration accuracy across various tasks and exemplifies vision transformer-based registration methods. 
It benefits from Swin Transformer's extensive effective receptive field \cite{luo2016understanding}. 
TransMorph has four variants, each utilizing a different Swin backbone. 
Our experiments utilized the 'Large' variant, which delivers the highest registration accuracy among its counterparts.

\vspace{1ex}
\noindent\textbf{LapIRN \cite{mok2020large,mok2021large}.}
LapIRN, based on residual ConvNets \cite{he2016deep}, features a three-level coarse-to-fine architecture.
Unique among its peers, LapIRN employs three sub-networks, each tailored to a specific scale of deformation, making it particularly effective for datasets with large deformations. 
The network's complexity is managed through its initial channel count, and for consistency with our textSCF framework, we set this count to 32 to get optimal performance.

\vspace{1ex}
\noindent\textbf{LKU-Net \cite{jia2022u}.}
LKU-Net, leveraging large kernel insights \cite{ding2022scaling} in a U-Net \cite{ronneberger2015u} framework, excels in producing both fine-grained fields and those with large deformation.
It was the chosen backbone for constructing the textSCF framework. 
The complexity of LKU-Net, like other models in our study, is regulated by the start channel count; we used 32 to maintain consistency with the textSCF setup.

\vspace{1ex}
\noindent\textbf{Fourier-Net \cite{jia2023fourier}.}
Fourier-Net generates deformation fields in a low-frequency space, simplifying network complexity while enhancing field smoothness. 
It claims to match or surpass the registration accuracy of TransMorph and LapIRN in brain registration tasks, but with reduced computational demands. 
However, its performance dips in datasets characterized by large deformations. 
In our implementation of Fourier-Net, we set the starting channel count to 32 and adjusted the low-frequency patch size to be a quarter of the original size, aligning with its most extensive variant.

\section{Diffeomorphic Registration}
\label{sec:diff_reg}
In this section, we explore the principles of diffeomorphic registration.
We begin by defining a diffeomorphism and discussing its relevance to the smoothness of deformation fields. 
We then present the concept of the Jacobian determinant as it pertains to deformation fields and introduce the formula for calculating the SDlogJ metric. 
Finally, we discuss the statistical underpinnings that allow SDlogJ to serve as an indicator of the quality of a diffeomorphic transformation.

\subsection{Diffeomorphism}

\noindent\textbf{Diffeomorphism Definition.}
In mathematics, a diffeomorphism represents an isomorphism between smooth manifolds. 
This entails an invertible function that connects one differentiable manifold to another, ensuring that both the function and its inverse are smooth and continuously differentiable.

\vspace{1ex}
\noindent\textbf{Diffeomorphism in Image Registration.}
In the context of deformable image registration, a diffeomorphism refers to a smooth and invertible transformation process. 
This allows for seamless image transitions without tearing or folding, ensuring the preservation of topologies. 
While a continuously differentiable transformation and its inverse typically define a diffeomorphism, we emphasize smoothness as a crucial and more manageable aspect within learning frameworks. 
To promote this smoothness, global regularizers \cite{balakrishnan2019voxelmorph} or a diffeomorphic integration layer are often employed \cite{ashburner2007fast,dalca2018unsupervised}. 
However, prioritizing smoothness might affect the accuracy of anatomical correspondences. Thus, striking a balance between smooth transformation and accurate anatomical mapping is crucial.

\subsection{Diffeomorphic Integration Layer}
Adding a differentiable diffeomorphic integration layer \cite{dalca2019unsupervised} with a global smoothness regularizer improves the smoothness of the deformation field. 
This process involves integrating a stationary velocity field \( \mathbf{u} \) over time $t=[0,1]$ to compute the final registration field \( \phi^{(1)} \). 
Beginning with the identity transformation \( \phi^{(0)} = Id \), the deformation field evolves according to \( \frac{\partial \phi^{(t)}}{\partial t} = \mathbf{u}(\phi^{(t)}) \). 
Stationary ordinary differential equation integration represents a one-parameter subgroup of diffeomorphisms. 
In group theory, \( \mathbf{u} \), part of the Lie algebra, is exponentiated to yield \( \phi^{(1)} = \exp(\mathbf{u}) \), a Lie group member. 
One-parameter subgroups ensure \( \exp((t + t')\mathbf{u}) = \exp(t\mathbf{u})~\circ~\exp(t'\mathbf{u}) \) for any scalars \( t \) and \( t' \), with \( \circ \) being the composition in the Lie group. 
Starting with \( \phi^{(1/2^T)} = p + \mathbf{u}(p) \), we use \( \phi^{(1/2^{t+1})} = \phi^{(1/2^t)} \circ \phi^{(1/2^t)} \) to get \( \phi^{(1)} = \phi^{(1/2)} \circ \phi^{(1/2)} \). 
In neural networks, diffeomorphic integration employs spatial transformation \cite{jaderberg2015spatial} layers for scaling and squaring operations \( T \) times. 
Based on \cite{dalca2018anatomical}, \( T=7 \) is our chosen number of integration steps.

\subsection{Smoothness Measurement: SDLogJ}
As previously noted, while a sufficiently smooth deformation field doesn't automatically ensure diffeomorphism, it is a crucial component in achieving it. 
The Jacobian matrix, formed by the deformation field's derivatives in each direction, plays a key role in understanding the deformation's local behavior. 
This matrix is essentially a second-order tensor field that captures these local changes.
The definition of the Jacobian matrix $J_{\phi}(p)$ is as follows:
\begin{equation}
J_{\phi}(p) = \left( 
\begin{array}{ccc}
\frac{\partial \phi_x(p)}{\partial x} & \frac{\partial \phi_x(p)}{\partial y} & \frac{\partial \phi_x(p)}{\partial z} \\
\frac{\partial \phi_y(p)}{\partial x} & \frac{\partial \phi_y(p)}{\partial y} & \frac{\partial \phi_y(p)}{\partial z} \\
\frac{\partial \phi_z(p)}{\partial x} & \frac{\partial \phi_z(p)}{\partial y} & \frac{\partial \phi_z(p)}{\partial z} \\
\end{array} 
\right),
\label{eq:jacobian}
\end{equation}
where $p$ is the voxel position, $\phi$ is the deformation field.
The Jocobian determinant of the deformation field at position $p$, denoted as $|J_{\phi}(p)|$, is useful in analyzing the local characteristics of the deformation field. 
In regions where the Jacobian determinant is positive, the local deformation field typically exhibits diffeomorphic properties, indicating a one-to-one mapping. 
Conversely, areas with a negative Jacobian determinant suggest a loss of this one-to-one correspondence, highlighting areas of concern in the deformation process.

\vspace{1ex}
\noindent\textbf{Interpretation of $| J_{\phi}(p) |$ values.}
When \( | J_{\phi}(p) | > 1 \), it corresponds to local expansion, indicating an increase in the volume at position $p$.
A determinant between 0 and 1, \( 0 < | J_{\phi}(p) | < 1 \), corresponds to local contraction, reflecting a reduction in volume.
If \( | J_{\phi}(p) | = 1 \), the deformation maintains the region's original size, reflecting volume preservation.
A determinant of zero, \( | J_{\phi}(p) | = 0 \), is associated with a collapse to a lower-dimensional structure, often a folding or singularity.
Negative values of the determinant, \( | J_{\phi}(p) | < 0 \), are associated with local inversion, where the orientation at that position is reversed, typically considered a non-physical transformation.

\vspace{1ex}
\noindent\textbf{Derivation of SDlogJ.}
SDlogJ serves as a valuable statistical metric \cite{leow2007statistical} for assessing the diffeomorphic properties of deformation fields in image registration. 
It quantitatively evaluates the uniformity and smoothness of deformations across the image, aiding in determining their physical viability and alignment with diffeomorphic characteristics.
Let $\mu$ be the mean of log Jacobian determinants and $N$ the total number of voxel positions in the field. 
The standard deviation of these determinants is defined as:
\begin{equation}
    \text{SDlogJ} = \sqrt{\frac{1}{N-1} \sum_{p} (\log \sigma(| J_{\phi}(p) |+\rho) - \mu)^2},
\end{equation}
where $\sigma$ is a clip function ensuring positive values and $\rho$ offsets highly negative values (following the challenge host \cite{hering2022learn2reg}, we set $\rho=3$).
SDlogJ statistically quantifies uniformity in a deformation field's transformation properties. 
A lower SDlogJ indicates a smoother, higher quality field with consistent local transformations. 
In contrast, a higher SDlogJ points to variable, potentially less smooth transformations.

\subsection{Diffeomorphic Quality Measurement: SDLogJ}

The significance of the Jacobian determinant in analyzing local deformation behavior at specific positions was previously emphasized. 
A non-positive Jacobian determinant indicates a disruption in bijective mapping at those locations, detracting from the field's diffeomorphic quality. 
Consequently, calculating the proportion of positions where \( J_{\phi}(p) \leq 0 \), represented as \( |J_{\phi}|_{\leq 0}\% \), provides a valuable metric for assessing the overall quality of diffeomorphism \cite{jia2023fourier,chen2022transmorph,balakrishnan2019voxelmorph}.

\begin{figure}[!t]
    \centering
    \includegraphics[width=0.9\columnwidth]{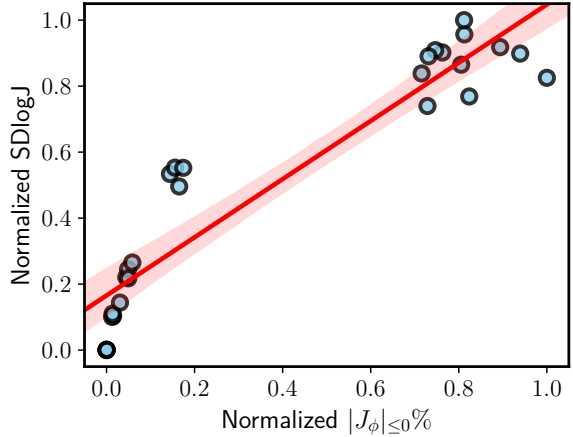}
    \caption{
        Scatter plot illustrating the relationship between normalized SDlogJ and normalized \( |J_{\phi}|_{\leq 0}\% \). Each point represents an observation from the textSCF models trained on the Abdomen dataset. The linear regression line, along with its confidence interval, suggests a strong positive correlation between the smoothness of the deformation field (SDlogJ) and the proportion of non-positive Jacobian determinants (\( |J_{\phi}|_{\leq 0}\% \)), reinforcing SDlogJ's role as a reliable indicator of diffeomorphic quality.
    } 
    \label{fig:jdet_sdlogj}
    \vspace{-2ex}
\end{figure}

\vspace{1ex}
\noindent\textbf{Experiment Settings.}
To examine the relationship between SDlogJ and \( |J_{\phi}|_{\leq 0}\% \), we conducted experiments using textSCF on the Abdomen dataset. 
Two factors influencing the smoothness of the generated deformation field were considered: the use of the diffeomorphic integration layer and the coefficient \( \lambda \) of the global smoothness regularizer. 
The following settings were employed to create a range of SDlogJ-\(|J_{\phi}|_{\leq 0}\%\) pairs: 1) \textit{External Segmentor}: Different external segmentation masks used in textSCF variants.
2) \textit{Network Complexity}: Variation in the starting channel count \( N_s \) from 8 to 16, and then to 32.
3) \textit{Diffeomorphic Integration}: Variants with and without the diffeomorphic integration layer.
4) \textit{\( \lambda \) for Smoothness Regularizer}: Adjusting \( \lambda \) from 0.01, 0.05, 0.1, to 1.0.

\vspace{1ex}
\noindent\textbf{Results and Analysis.}
By adjusting the mentioned variables, we trained various textSCF models, generating a range of SDlogJ-\(|J_{\phi}|_{\leq 0}\%\) values. 
The results showed that SDlogJ values ranged between $(6.75e-02, 9.50e-01)$ and \( |J_{\phi}|_{\leq 0}\% \) values between $(3.37e-06, 4.16e-02)$.
To improve clarity, min-max normalization was applied to both SDlogJ and \( |J_{\phi}|_{\leq 0}\% \). 
This scaling brings their values within the $(0,1)$ range, ensuring data consistency and improved readability, while preserving the correlation between the two variables.
Fig. \ref{fig:jdet_sdlogj} displays scatter plots of normalized SDlogJ-\(|J_{\phi}|_{\leq 0}\%\) with linear regression analysis. 
The Pearson correlation coefficient of 0.93 ($p<0.05$) indicates a strong positive correlation, showing that increases in SDlogJ typically accompany rises in \(|J_{\phi}|_{\leq 0}\%\) and vice versa. 
An R-squared value of 0.86 in our regression model reveals that SDlogJ explains about 86\% of the variability in \(|J_{\phi}|_{\leq 0}\%\). 
This suggests SDlogJ's high predictive value for \(|J_{\phi}|_{\leq 0}\%\), indicating its effectiveness as a substitute metric. 
Thus, SDlogJ measurements can effectively parallel insights gained from \(|J_{\phi}|_{\leq 0}\%\) evaluations.
Combined with SDlogJ's capability to assess deformation field smoothness, it stands as a straightforward metric for evaluating diffeomorphism quality.

\vfill

\end{document}